\documentclass[10pt,twocolumn]{revtex4}
\usepackage{graphicx}
\usepackage{psfig}
\newcommand{\be}{\begin{equation}}
\newcommand{\bea}{\begin{eqnarray}}
\newcommand{\ee}{\end{equation}}
\newcommand{\eea}{\end{eqnarray}}

\begin{document}
\title{Integrability, stability, and adiabaticity in nonlinear stimulated Raman adiabatic passage.}

\author{A.P. Itin$^{1,2}$, S. Watanabe$^1$.} \affiliation{$^1$University of
Electro-Communications, 1-5-1, Chofu-ga-oka, Chofu-shi, Tokyo
182-8585, Japan\\ $^2$Space Research Institute, RAS, Profsoyuznaya
str. 84/32, 117997 Moscow, Russia}

\begin{abstract}
We study dynamics of a two-color photoassociation of atoms into
diatomic molecules via nonlinear Stimulated Raman adiabatic
passage (STIRAP) process. This system has a famous counterpart in
(linear) quantum mechanics, and been discussed recently in the
context of generalizing quantum adiabatic theorem to nonlinear
systems. Here we use another approach to study adiabaticity and
stability in the system: we apply methods of classical Hamiltonian
dynamics. We found nonlinear dynamical instabilities, cases of
complete integrability, and improved conditions of adiabaticity.
\end{abstract}

\maketitle Adiabatic theorem \cite{Sanders} of quantum mechanics
has found wide applications in quantum state manipulations.
Dynamics of Bose-Einstein condensates (BEC) \cite{GP} introduces
paradigm of nonlinearity into a quantum world (note studies on
nonlinear Landau-Zener models \cite{Zobay,IW}, solitons, shock
waves, etc). Along with nonlinearity, it also brings a question of
how to analyze adiabaticity in nonlinear systems. It is not
justified to apply the adiabatic condition of quantum mechanics to
nonlinear BEC systems \cite{Pu,Reinhardt}. Instead, a general
method was recently suggested \cite{Pu} and applied to a specific
example of two-color Raman photoassociation system
\cite{Javanainen}. Here, we consider the same example as in
\cite{Pu,Javanainen} and notice that, as a typical nonlinear
system, it possesses many counterintuitive difficulties in the
analysis. The goal of our paper is twofold. Firstly, we (inspired
by ideas of \cite{Pu}) suggest another approach to analyze
adiabaticity and stability in nonlinear BEC-related systems: we
use methods of classical Hamiltonian dynamics \cite{AKN}.
Secondly, we improve a theory of nonlinear STIRAP (Stimulated
Raman adiabatic passage).
\begin{figure}
\includegraphics[width=75mm]{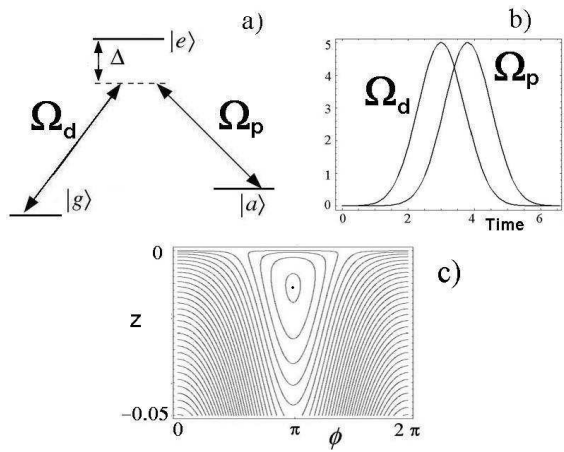}
\caption{(a) A three-level $\Lambda$-system  (b) Counterintuitive
STIRAP sequence of pulses $\Omega_{p,d}$ (c) Phase portrait of
(\ref{cham}).  As parameters are changing in accordance with a
pulse sequence shown in (b), the stable fixed point (corresponding
to the dark state) moves from $z=0$.} \label{Lambda}\end{figure}

Following \cite{Pu, Javanainen}, consider a three-level
$\Lambda$-system shown in Fig. \ref{Lambda}a. The state
$|a\rangle$ corresponds to an atomic BEC, which is coupled to an
excited diatomic molecular BEC (the state $|e\rangle$) via a Raman
laser pulse $\Omega_p$ ("pump"). A single-photon detuning $\Delta$
is the difference between a frequency of the pump laser and the
frequency of transition between $|a\rangle$ and $|e\rangle$. The
state $|e\rangle$ is, in turn, coupled to the ground state of the
molecular BEC by a laser pulse $\Omega_d$ ("dump"). In other
words, a pair of atoms from $|a\rangle$  is photoassociated by a
pump field of molecular Rabi frequency $\Omega_p$ into a molecule
in the excited state $|e\rangle$, which is subsequently driven
into the ground molecular state. This system is intrinsically
nonlinear, as it involves merging of atoms into diatomic
molecules.  Its linear counterpart is widely used in quantum
manipulations: STIRAP \cite{STIRAP} is well-established technique
for transferring population from $|a\rangle$ to $|g\rangle$ in
three-level quantum systems. The gist of the technique is as
following. Intuitively, one may think of transferring firstly the
population from $|a\rangle$ to $|e\rangle$ by means of $\Omega_p$
pulse, and then from $|e\rangle$ to $|g\rangle$. However, the
state $|e\rangle$ often suffers from spontaneous emission. STIRAP
counterintuitive sequence of pulses \cite{STIRAP} consists of
overlapping $\Omega_{p,d}$ pulses, with $\Omega_d$ going first
(see Fig. \ref{Lambda}b). STIRAP process in both linear and
nonlinear ${\Lambda}$-systems is based on the existence of a {\em
dark state} which is a superposition of only $|a\rangle$ and
$|g\rangle$ states (and therefore almost does not suffer from
spontaneous emission of light, which explains the notation;
another notation for the dark state is coherent population
trapping (CPT) state \cite{Pu}). An instantaneous dark state is
determined by values of $\Omega_{p,d}$; population transfer
happens via the dark state as $\Omega_{p,d}$ are slowly changed,
with $|e\rangle$ state remaining almost unpopulated. Stability and
adiabaticity of the system in the dark state is essential for
efficiency of population transfer. In contrast to its linear
counterpart, the theory of nonlinear STIRAP is not well developed
yet. This process is very important for BEC physics, as it allows
to achieve BEC of molecules \cite{Winkler}. Mean-field approach
has shown to be reliable in such systems
\cite{Mackie,Javanainen,Pu}. Neglecting mean-field collisional
interactions and spontaneous emission, the mean-field equations
are \cite{Pu}
\bea i \dot{\psi}_a &=& \Omega_p \psi_a^* \psi_e, \quad  i \dot{\psi}_g =  \frac{\Omega_d}{2} \psi_e, \nonumber\\
 i \dot{\psi}_e &=& \Delta \psi_e + \frac{ \Omega_p}{2} \psi_a^2 + \frac{\Omega_d}{2} \psi_g, \label{dyn}
\eea where amplitudes $\psi_{a,e,g}$ are normalized as $|\psi_a|^2
+ 2(|\psi_g|^2 + |\psi_e|^2) =1 $. The dark state vector is given
by (up to a phase factor) ${\bf \Psi}_0 =
(\psi_a^0,\psi_e^0,\psi_g^0)^T $, where $ \psi_a^0 = \Bigl[
\frac{2 \Omega_d}{\Omega_d + \Omega_e} \Bigr]^{1/2}, \psi_e^0 = 0,
\psi_g^0= - \frac{2 \Omega_p}{\Omega_d + \Omega_e}, $ with
$\Omega_e=\sqrt{\Omega_d^2+8\Omega_p^2}$. Linearization about the
dark state \cite{Pu} gives a dynamical system with three
eigenfrequencies: $ \omega_0=0, \quad \omega_{\pm} =
\frac{1}{2}[\Delta \pm (\Delta^2+ \Omega_d \Omega_e)^{1/2}].$ The
frequencies $\omega_{0,\pm}$ are all real. However, it does not
guarantee that the system is always dynamically stable: there
might be nonlinear instabilities.

We recast the model into the form of a two-degree-of-freedom (2
d.o.f.) classical Hamiltonian system, and use the classical
adiabatic theory and the resonance normal forms theory available
in \cite{AKN,Duist}. Eqs. (\ref{dyn}) are equivalent to
Hamiltonian equations of motion of the effective classical
Hamiltonian \be H=\Omega_p \left[x_2 \frac{y_1^2-x_1^2}{2}-x_1 y_2
y_1\right] - \frac{\Omega_d}{2}(x_2 x_3 +y_2 y_3) - \Delta
\frac{x_2^2+y_2^2}{2}. \label{xy} \ee Here $x_k$ are canonical
momenta, while $y_k$ are the coordinates, being related to the old
"variables" (complex numbers $\psi_i$) as $ \psi_a = x_1 + i y_1,
\quad \psi_e = x_2 + i y_2,  \quad \psi_g = x_3 + i y_3. $ The
system has two degrees of freedom only: there exists integral of
motion $N=x_1^2+y_1^2 + 2(x_2^2+y_2^2+x_3^2+y_3^2)=1$.

\begin{figure}
\includegraphics[width=80mm]{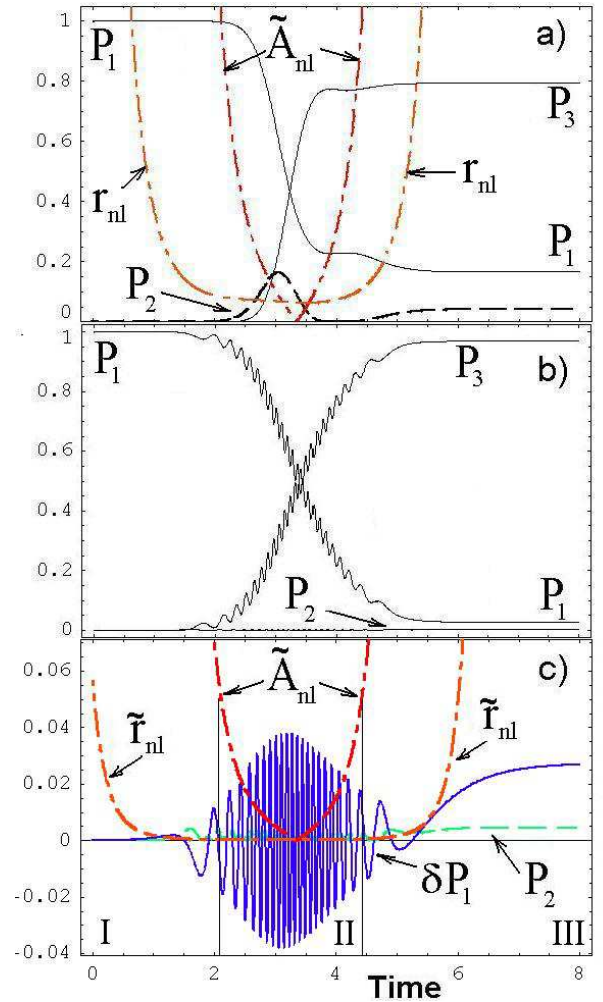}
\caption{(Color online)Population dynamics and the adiabatic
parameters $A_{nl}$, $r_{nl}$. Tildes over the parameters denote
division by 10 (in order to fit into the scale): $\tilde X \equiv
X/10$. $P_{1,2,3}$: populations of $|a\rangle$, $|e\rangle$,
$|g\rangle$, i.e. $x_1^2+y_1^2, 2(x_2^2+y_2^2), 2(x_3^2+y_3^2) $
correspondingly. Solid lines: $P_{1,3}$; dashed lines: $P_2$
(should be zero in the dark state); dot-dashed lines: the
adiabatic parameters. (a) The same numerical example as in
Ref.\cite{Pu}; $t_p=3.8, t_d=3$, $\Omega_0=5$. Note the dynamics
of $P_2$: at $t \approx 3$ it deviates from zero considerably.
Parameter $r_{nl}$ suggests adiabatic evolution at $1.5<t<4.5$.
Parameter $A_{nl}$ is huge (almost two orders of magnitude larger
than $r_{nl}$), suggesting nonadiabatic evolution. Note that
$H,I_{1,3}$ remains zero.  (b,c) $\Omega_0=100$, other parameters
are the same as in the previous numerical example; (b) Population
dynamics. High-frequency oscillations are seen in $P_{1,2,3}$ (c)
Dynamics of adiabatic parameters $A_{nl},r_{nl}$ and deviation of
$P_1$ from its value in the dark state ($P_1^0$): $\delta
P_1=P_1-P_1^0$. Three stages of nonlinear STIRAP can be defined
(I-III), with only the second stage being adiabatic. In the first
stage ($t\lesssim 2.5$) $A_{nl}$ is large, so system acquires some
action. At the second stage ($2.5\lesssim t\lesssim 4.5 $)
$A_{nl}$ is small, and population transfer proceeds (with
approximate conservation of the acquired action). At the third
stage (from $t \approx 4.5$), adiabaticity is broken again.
Predictions of $A_{nl}$ are better than that of $r_{nl}$. }
\label{Fpopulation}
\end{figure}

Firstly, we consider the case $\Delta=0$ (single-photon
resonance). Let $H=0$. On this manifold, dynamics at constant
parameters $\Omega_{p,d}$ is completely integrable. Indeed, $I_2
\equiv x_2/y_2$ is the additional integral of motion. As a result,
$I_3 \equiv I_2 x_3 + y_3$ and $I_1 \equiv I_2
\frac{y_1^2-x_1^2}{2} -x_1 y_1  $ are also integrals of motion.
Integrability of the system with $\Delta=0$ at $H=0$ energy
manifold is an important finding of this paper. In case
$\Omega_{p,d}$ are changing with time, the system may leave the
$H=0$ energy manifold. However, in case initial conditions are
such that $I_{1,3}=0$, then even with time-dependent parameters
dynamics will be confined to the initial $H=0$ manifold.  The
particular case considered in \cite{Pu} (where all population is
initially in $|a\rangle$) is of this type. Indeed, initially
$x_{2,3}=y_{2,3}=0$ (therefore, $H=0$). As parameters are changed,
$I_1,I_3$ and $H$ all remain equal to zero, see Fig.
\ref{Fpopulation}, where the example of \cite{Pu} is presented.
The pulses are $ \Omega_{p,d}=\Omega_0 e^{-(t-t_{p,d})^2},$ with
$t_d<t_p$. With $H=\mbox{const} \equiv 0$, we can reduce the
system to a $1 \frac{1}{2}$ d.o.f. classical Hamiltonian. Without
loss of generality, let $I_2=0$. Therefore, only $x_1,x_3,y_2$ are
non-zero during the evolution. The equations of motion are $
\dot{x_1} = \Omega_p x_1 y_2, \quad
   \dot{x_3} = \frac{\Omega_d}{2} y_2, \quad
   \dot{y_2} = -\frac{\Omega_d}{2} x_3 - \frac{\Omega_p}{2} x_1^2.
$ Introducing new variables $\phi$ and $z$ as
$x_3=\frac{1}{\sqrt{2}}\sqrt{1-x_1^2} \cos \phi, \quad
y_2=\frac{1}{\sqrt{2}}\sqrt{1-x_1^2} \sin \phi, \quad
 x_1=\exp[z],$ the equations of motion correspond now to the classical Hamiltonian \be {\cal H}=
-\frac{\Omega_d}{2} z + \frac{\Omega_p}{\sqrt{2}} \sqrt{1-e^{2z}}
\cos \phi, \label{cham} \ee where $\phi$ and $z$ are canonically
conjugated variables.
Fig. \ref{Lambda}c presents phase portraits of this Hamiltonian.
There is a stable fixed point which corresponds to the dark state:
$z= \frac{1}{2}\ln \large[ \frac{2 \Omega_d}{\Omega_d + \Omega_e}
\large], \quad \phi=\pi $.  As time increases, it goes from $z=0$
to $-\infty$. For small-amplitude oscillations, linearization
around the stable fixed point gives the frequency $
\omega=\frac{1}{2} (\Omega_d \Omega_e)^{1/2}.$ Adiabatic evolution
requires change of this frequency $\delta \omega $ at one period
of unperturbed motion to be much less than frequency itself
\cite{AKN,how}: $|\delta \omega| \ll |\omega|$. Since $ \delta
\omega \approx \dot{\omega}* 2\pi/\omega $, we have the following
criteria: $|\dot{\omega}| \ll \omega^2/ 2 \pi$, or $ |\Omega_e
\dot{ \Omega}_d + \dot{\Omega}_e \Omega_d| \ll
 \frac{1}{2 \pi}(\Omega_d \Omega_e)^{3/2}, $
or \be A_{nl} \equiv 2 \pi | \Omega_e \dot{ \Omega}_d +
\dot{\Omega}_e \Omega_d|/(\Omega_d \Omega_e)^{3/2} \ll 1
\label{criterion} \ee It can be seen from Fig. \ref{Fpopulation}
that $A_{nl}$ works sufficiently better than $r_{nl}$ of \cite{Pu}
($A_{nl}$ is typically an order of magnitude larger than $r_{nl}$
in this numerical example). For one-and-half d.o.f. Hamiltonian
systems, the criterion $|\delta \omega| \ll |\omega|$ as described
above is reliable to analyze adiabaticity.

In systems with several degrees of freedom, situation is much more
complicated. Even stability at fixed parameters is highly
nontrivial issue. Indeed, let us now consider the case $\Delta \ne
0$ (or $\Delta=0$, but $H \ne 0$), where we cannot utilize the
trick with the energy manifold, so the system with fixed
parameters $\Omega_{p,d}$ has 2 d.o.f. We make several canonical
transformations to reduce the Hamiltonian system to a 2 d.o.f.
one, and shift the origin to the point corresponding to the dark
state (see \cite{AV}). We get the Hamiltonian in new variables
$q_{1,2}, p_{1,2}$ \be {\cal H} = \Omega_p [q_1^2 + p_1^2 +q_2^2 +
p_2^2]p_1 - \frac{\Omega_d}{2}q_1 q_2- \frac{\Omega_e}{2} p_1 p_2
-  \frac{\Delta}{2}(q_1^2 + p_1^2). \label{qpA} \ee One may
transform the quadratic part of the Hamiltonian to its normal
form: sum of two linear oscillators, and neglect the cubic part,
obtaining therefore two uncoupled oscillators (with frequencies
$\omega_{\pm}$). However, such straightforward approach is
dangerous at low-order resonances between $\omega_{\pm}$.  The 1:1
resonance happens at $\Delta=0$, while 1:2 resonance at
$\Delta=\sqrt{\Omega_d \Omega_e/8}$. Consider firstly 1:1
resonance ($\Delta=0$). We introduce a parameter $\theta$ as $
\Omega_d = \Omega_e \cos \theta , \quad \Omega_p =  \Omega_e  \sin
\theta/2 \sqrt{2}, \quad 1> \cos \theta
>0, $ and divide the Hamiltonian (\ref{qpA}) by $\Omega_e$: $
{\cal H} = \frac{ \sin \theta}{2 \sqrt{2}} (q_1^2 + p_1^2 +q_2^2 +
p_2^2)p_1 - \frac{\cos \theta}{2} q_1 q_2 - \frac{1}{2} p_1 p_2. $

We transform the quadratic part of the Hamiltonian to the normal
form by means of linear transformations and obtain (retaining old
notations for new variables) \be H= H_2+H_3= \frac{1}{2}
\sqrt{\cos \theta} (p_2 q_1 - p_1 q_2 ) + H_3, \label{normal} \ee
where $H_3$ consists of cubic terms (for 1:1 resonance, it is
convenient to use series in $p_2 q_1 - p_1 q_2$, $p_1^2 + p_2^2$,
$q_1^2+q_2^2$. Hamiltonians which depends only on such
combinations of variables are also called normal forms). We need
to get rid of the cubic terms in the Hamiltonian. To this end, we
fulfill a nonlinear canonical transformation using a generating
function $K_3$ which is a homogenous polynomial of the third order
in old coordinates and new momenta: $ K_3= \alpha q_1^2 q_2 +
\beta P_2^2 q_2 + z P_1^2 q_2 + \gamma q_2^3 + f P_1 P_2 q_1.$ For
the coefficients $\alpha,\beta,\gamma,f,z$ in $K_3$, we get a
system of linear equations; the transformation defined by $K_3$
kills all cubic terms, while quadratic part remains the same; at
the same time, quartic terms emerge: we get the Hamiltonian $ H=
\frac{1}{2} \sqrt{\cos \theta} (P_2 Q_1 - P_1 Q_2 ) + H_4 + .., $
where $H_4$ contains the quartic terms. Generally, it is not
possible to kill all quartic terms \cite{AKN}. We need to
transform $H_4$ to the normal form. However, we can proceed in a
simpler (but equivalent) way. \begin{figure}
\includegraphics[width=72mm]{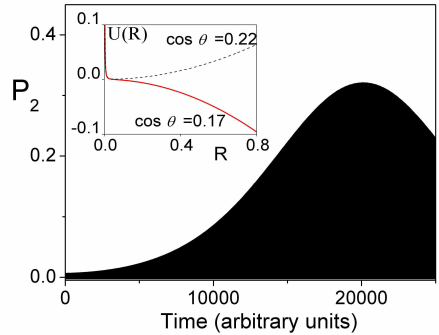}
\caption{(Color online) Instability in 1:1 resonance. Dynamics of
population of the excited state $P_2$ (which should be zero in the
dark state) is shown. Parameters: $\Omega_p=2.0$, $\Omega_d=1.0$
($\cos \theta \approx 0.17 < 0.2$, $\Omega_p> \sqrt{3} \Omega_d$).
Instability develops on long timescales: we start with small
deviation from the dark state, but the deviation grows and reach
very large values. With $\cos \theta
> 0.2$ (not shown), initial deviation do not grow much: it slowly oscillates about the equilibrium of the effective
potential of (7). The curve $P_2(t)$ is filled by oscillations
with frequency $\frac{1}{2}\sqrt{\Omega_d \Omega_e}$ which are not
seen in this scale. Inset shows two effective potentials of
(\ref{effective}) at $J =0.001$ and two different values of
$\theta$; solid line with $\cos \theta =0.17 <0.2$ corresponds to
the unbounded motion (instability).}
 \label{11resonance}
\end{figure} We change to polar coordinates in
$Q_1,Q_2$ plane making a transformation $ Q_1 = R \cos \xi, \quad
Q_2= R \sin \xi, \quad P_1 = P \cos \xi - \frac{J \sin \xi}{R},
\quad P_2 = P \sin \xi + \frac{J \cos \xi}{R},$ and then average
over the (fast) angle $\xi$. We obtain the following effective
Hamiltonian: \be F = \frac{1}{2} J - \frac{J \kappa_{\theta}^2}{8}
\left[ \frac{A_P}{2} \large( P^2 + \frac{J^2}{R^2} \large)   + R^2
A_R \right], \label{effective} \ee where $ A_P = \frac{10 \cos^2
\theta +10 \cos \theta +4}{3}, \quad A_R = 5 \cos \theta-1$,
$\kappa_{\theta} =\frac{\sin \theta}{ \sqrt{2} \sqrt{\cos
\theta}}$. In this Hamiltonian, $J$ is constant. Therefore, it is
an integrable system for a pair of canonically conjugated
variables $P,R$. It is not difficult to understand its dynamics:
(\ref{effective}) is merely a Hamiltonian of a particle in the
potential $U(R)= \frac{J^2 A_P}{2R^2 } + A_R R^2$ (indeed, only
the expression in the square brackets in (\ref{effective}) is
important for dynamics). The effective potentials $U(R)$ are shown
in the inset of Fig.(\ref{11resonance}). For $\cos \theta
> \frac{1}{5}$, the system (\ref{effective}) has a single fixed
point, while for $\cos \theta < \frac{1}{5}$ there are no fixed
points. In the latter case (corresponding to $\Omega_p/\Omega_d
>\sqrt{3}$), a phase point initially placed close to $R=0$ will slowly move from the origin to large values of $R$.
 The dark state corresponds to $R=0$, therefore this case implies dynamical instability due to 1:1 resonance. The
instability is slow (see Fig.\ref{11resonance}), nevertheless it
is physically important: it develops on the same timescales as
that used in \cite{Javanainen}. We checked that crossing the
critical $\cos \theta = \frac{1}{5}$ value results in onset of
dynamical instability, in accordance with the effective potential
(\ref{effective}). Physically, $\cos \theta <\frac{1}{5}$ means
that most of the population is in $|g\rangle$ state. In this
region, the dark state is unstable: deviations from it slowly
grows with time, populating $|e\rangle$ state considerably, which
would lead to spontaneous emission in real experiments. To be more
precise, deviations from the dark state firstly oscillate in time
with frequency $\frac{1}{2} \sqrt{\Omega_d \Omega_e} $ and small
amplitude, and then the amplitude of oscillations slowly grows.
Thus, one may see periodic bursts of spontaneous emission each
time the deviations reach their maxima (i.e., with a time period
$T \sim 1/\sqrt{\Omega_d \Omega_e}$).

Let us now turn to 1:2 resonance. Usually, its analysis is much
simpler than that of 1:1 and 1:3 resonances, as it involves only
cubic terms. Unfortunately, in our case  this resonance is also
degenerate. In 1:2 resonance, we can bring the Hamiltonian to the
resonance normal form \be H_{res} = \omega_1 R_1+ \omega_2 R_2 +
\kappa R_1^{1/2} R_2 \sin( \phi_1 + 2 \phi_2), \label{resform} \ee
where $(R_i,\phi_i)$ are symplectic polar coordinates
(action-angle variables of $H_2$: $P_i= \sqrt{2 R_i} \sin \phi_i,
\quad Q_i= \sqrt{2 R_i }\cos \phi_i $), $\kappa$ is some
coefficient, the frequencies $\omega_1$ and $\omega_2$ are in  1:2
resonance: $\omega_1 + 2\omega_2=0$ (so, $\omega_1 \omega_2<0$).
The resonance normal form is obtained by averaging out all terms
in $H_3$ except the resonance one (which depends on the resonance
phase $\gamma \equiv \phi_1+2 \phi_2$) \cite{AKN}. The resonance
normal form is integrable. Provided certain conditions of
generality are fulfilled, the equilibrium of the original system
(i.e., the dark state) is stable or unstable simultaneously with
the equilibrium of the resonance normal form (i.e., $R_{1,2}=0$).
For 1:2 resonance, a condition of generality is $\kappa \ne 0$;
with $\omega_1 \omega_2 <0$ and $\kappa \ne 0$, the equilibrium is
unstable \cite{AKN}. In our case, $\kappa=0$, therefore this
source of instability is absent in the system. Still, the
degeneracy $\kappa=0$ does not guarantee stability, but designates
that possible instabilities, if exist, are very slow.

It is important to emphasize that stability and adiabaticity are
two different issues. Adiabatic invariance in systems with several
degrees of freedom is also remarkably nontrivial \cite{Rbil}. Main
difficulties come from {\em passage through resonances}; in case
our system stays away from low-order resonances as parameters are
changed, we can consider it as two decoupled oscillators and
generalize the adiabatic condition (\ref{criterion})
straightforwardly, see \cite{AV}.

Here, we analyzed low-order resonances of the nonlinear $\Lambda$
system; found nonlinear instabilities of the dark state; found
cases of complete integrability and improved adiabatic conditions
for nonlinear STIRAP. The suggested method is generalized to
systems with mean-field collisional interactions in \cite{AV}.
Briefly, theory of nonlinear STIRAP is developed.

A.P.I. was supported by JSPS and 21st Century COE program on
``Coherent Optical Science''. This work was also supported by
Grants-in-Aid No. 16-04315 from MEXT, Japan. A.P.I. acknowledges
help of A.A.Vasiliev, discussions with A.I. Neishtadt, and thanks
J.Bohn for his kind invitation to visit JILA, where this
manuscript was finished.


\begin{thebibliography}{10}
\bibitem{how} We estimate influence of movement of the fixed point to
be of the same order as change in the frequency; the adiabatic
condition is therefore the same as in the case of a stationary
fixed point; generally, it is not always so.
\end{thebibliography}
\end{document}